\documentclass[aps,preprint,showpacs,floatfix]{revtex4}
\usepackage{graphicx,bm,amsmath,dcolumn,}
\usepackage{verbatim}
\usepackage{color}
\usepackage{epstopdf}
\begin{document}
\newcommand{\be}{\begin{equation}}
\newcommand{\ee}{\end{equation}}
\newcommand{\half}{\frac{1}{2}}
\newcommand{\ith}{^{(i)}}
\newcommand{\im}{^{(i-1)}}
\newcommand{\gae}
{\,\hbox{\lower0.5ex\hbox{$\sim$}\llap{\raise0.5ex\hbox{$>$}}}\,}
\newcommand{\lae}
{\,\hbox{\lower0.5ex\hbox{$\sim$}\llap{\raise0.5ex\hbox{$<$}}}\,}

\definecolor{blue}{rgb}{0,0,1}
\definecolor{red}{rgb}{1,0,0}
\definecolor{green}{rgb}{0,1,0}
\newcommand{\blue}[1]{\textcolor{blue}{#1}}
\newcommand{\red}[1]{\textcolor{red}{#1}}
\newcommand{\green}[1]{\textcolor{green}{#1}}

\newcommand{\yc}{y_{\rm can}}

\newcommand{\calA}{ {\mathcal A}}
\newcommand{\calC}{ {\mathcal C}}
\newcommand{\calE}{ {\mathcal E}}
\newcommand{\calG}{ {\mathcal G}}
\newcommand{\calH}{ {\mathcal H}}
\newcommand{\calO}{ {\mathcal O}}
\newcommand{\calS}{ {\mathcal S}}
\newcommand{\calZ}{ {\mathcal Z}}

\newcommand{\arho}{ {\overline{\rho}}}
\title{Percolation in the canonical ensemble}
\author{Hao Hu$^1$, Henk W. J. Bl\"ote$^{2}$ 
and Youjin Deng$^1$\footnote{yjdeng@ustc.edu.cn}}
\affiliation{$^{1}$ Hefei National Laboratory for Physical
Sciences at Microscale,
Department of Modern Physics, University of Science and
Technology of China, Hefei 230027, China }
\affiliation{$^{2}$ Instituut Lorentz, Leiden University,
P.O. Box 9506, 2300 RA Leiden, The Netherlands}
\date{\today}

\begin{abstract}
We study the bond percolation problem under the constraint that
the total number of occupied bonds is fixed, so that the canonical ensemble applies.
We show via an analytical approach that at criticality, the constraint can induce new 
finite-size corrections with exponent $\yc=2y_t-d$ both in energy-like and magnetic quantities, 
where $y_t=1/\nu$ is the thermal renormalization exponent and $d$ is the spatial dimension.
Furthermore, we find that while most of universal parameters remain unchanged, 
some universal amplitudes, like the excess cluster number, can be modified and become non-universal.
We confirm these predictions by extensive Monte Carlo simulations of 
the two-dimensional percolation problem which has $\yc=-1/2$.
\end{abstract}
\maketitle 

\section{Introduction}
\label{intro}
The bond-percolation model \cite{SA} can be considered as the $q \rightarrow 1$
limit of the $q$-state Potts model \cite{RBP,FW}.  Consider a lattice 
$G \equiv (V,E)$ with $V$ ($E$) the vertex (edge) set, the reduced (divided
by $kT$) Hamiltonian of the Potts model reads
\begin{equation}
  \calH (K,q) = -  K \sum_{ij \in E } \delta_{\sigma_i \sigma_j}  
  \hspace{10mm} (\sigma = 1,\cdots, q) \; ,
\label{Potts}
\end{equation}
where $K$ is the coupling strength.
By introducing bond variables on the edges and integrating out the spin degrees of freedom, 
one can map the Potts model onto the random-cluster (RC) model.
This is the celebrated Kasteleyn-Fortuin transformation~\cite{KF}.  
The partition sum of the RC model thus assumes the form
\begin{equation}
\calZ_{\rm rc}(u,q) = \sum_{\calA \subseteq G} \, u^{|\calA|} q^{k(\calA)} 
\hspace{10mm} ( u= e^K-1) \; , 
\label{zrc}
\end{equation}
where the sum is over all the spanning subgraphs $\calA$ of $G$, $|\calA|$ is the
number of occupied bonds in $\calA$, and $k (\calA)$ is the number of connected 
components (clusters).  The Kasteleyn-Fortuin mapping provides a 
generalization of the Potts model to non-integer values $q >0$. 
In the $q \to 1$ limit, the RC model reduces to the bond 
percolation problem \cite{KF,FW}. In this limit, the partition sum assumes
the non-singular form $(u+1)^{|E|}$, where $|E|$ is the number of the lattice
edges. Nevertheless, rich physics still exists, e.g. in the derivatives of the
partition sum involving $q$.

Percolation in two dimensions has been extensively studied. 
Percolation thresholds on many two-dimensional 
lattices are exactly known or have been determined to very high precision
(see e.g. Ref.~\cite{Ziff11} and references therein).  
Most of the critical exponents are also exactly known. 
For instance, the leading and subleading thermal renormalization exponents 
are $y_t =1/\nu = 3/4$ and $y_i = -2$, respectively; the leading two critical exponents 
in the magnetic sector are $y_h = d-\beta/\nu = 91/48$ and $y_{h2} = 71/96$ \cite{BNDL, JCDL}. 
There still exist some critical exponents whose exact values are unknown, 
including the backbone exponent and the shortest-path fractal 
dimension \cite{GB, ZYDZ}. Much progress has recently been made in the context 
of the stochastic-L\"owner evolution \cite{WKBN, JC}.

The present work provides a study of another aspect of percolation. 
We introduce an ``energy-like'' constraint that the total number 
of occupied bonds is fixed and study the effects of such a constraint on the 
critical finite-size scaling (FSS).  Since occupied bonds can formally be
considered as particles, we shall refer to percolation 
under this constraint as percolation in the canonical ensemble,
and to the case without the constraint as percolation in the
grand-canonical ensemble. Two of us have, for several years,
been studying  the effects of such constraints \cite{DB, DHB, DB1}.  
The leading FSS in systems under energy-like constraints is derived on
the basis of the so-called Fisher renormalization~\cite{MF} procedure, 
which was originally formulated for systems in the thermodynamic limit. 
Consider an energy-density-like observable $\varepsilon$ in the critical RC model
that scales as $ \langle \varepsilon (L) \rangle =\varepsilon_a +\varepsilon_s L^{y_t-d} $ 
in the grand-canonical ensemble, where $L$ is the linear system size and
$d$ is the spatial dimensionality. 
The constraint in the canonical ensemble implies: \\
(i), $\varepsilon (L) $ for a given size $L$ is fixed at the  expectation
value $\varepsilon_a$ in the thermodynamic limit. This expectation value
is normally different from the grand-canonical finite-size average 
$ \langle \varepsilon (L) \rangle$. \\ 
(ii), fluctuations of $\varepsilon$ are forbidden---i.e.,
$\langle \varepsilon^2 \rangle = \langle \varepsilon \rangle^2$.  \\
The effect of (i) is accounted for  by the Fisher renormalization. 
The bond density $|\calA|/|E|$ is such an observable. 
For bond percolation, however, since the bond variables on different edges 
are independent of each other, 
$\rho_b=\langle |\calA| \rangle/|E|$  does not depend on the system size
$L$---i.e., if we write $\rho_b=\rho_{b,a}+ \rho_{b,s} L^{y_t-d}$, then 
the amplitude $\rho_{b,s}=0$.
Thus, percolation provides an 
ideal system to study the effects of energy-like constraints due to 
the suppression of fluctuations.

The remainder of this work is organized as follows. 
Section II describes the sampled quantities and the FSS of physical observables for percolation 
in the grand-canonical ensemble. 
Section III derives the effects of the ``energy-like'' constraint in the critical FSS. 
The numerical results are presented in Sec.~\ref{num_results}. 
A brief discussion is given in Sec.~\ref{discus}. 

\section{Sampled quantities and finite-size scaling}
\label{simul}
\subsection{Simulation and sampled quantities}
We consider the bond-percolation problem on an $L \times L$ square lattice with 
periodic boundary conditions.  The grand-canonical simulations follow the
standard procedure: each edge is occupied by a bond with probability $p$,
after which the percolation clusters are constructed.  For simulations in 
the canonical ensemble, a Kawasaki-like scheme \cite{Kawasaki} is used.  
Given an initial configuration with a total number of occupied bonds
$|\calA|=p |E|$, where $|E|=2 L^2$, an update is defined as the random 
selection of two edges and the exchange of their occupation states.  
Each sweep consists of $2 L^2 $ updates.  Quantities are sampled after every sweep.  
Note that although the bond number $|\calA|=p |E|$ happens to be an integer 
for the critical square-lattice bond percolation, $p |E|$ is generally a real number.
In this case, one can simulate at the ceiling and the floor integer of $p |E|$ 
and apply an interpolation.

Given a configuration $\calA$ as it occurs in the simulation, we denote 
the sizes of clusters as $\calC_i$ ($i=1,\ldots,k (\calA)$); 
$\calC_1$ is reserved for the largest cluster. 
The following observables were sampled.
\begin{enumerate}
\item {\it Energy-like quantities}: 
  \begin{itemize}
    \item The bond-occupation density $\rho _b =\langle |\calA| \rangle/|E|$. 
    \item The cluster-number density $\rho _k =\langle k(\calA) \rangle/|V|$. 
  \end{itemize}

\item {\it Specific-heat-like quantities}: 
  \begin{itemize}
    \item $C_b = 
    L^{-d} \left(\langle |\calA|^2\rangle - \langle |\calA|\rangle^2 \right)$.
    \item $C_k = 
    L^{-d} \left(\langle k(\calA)^2 \rangle - \langle k(\calA) \rangle^2 \right)$.
    \item $C_{kb} 
    = L^{-d} \left(\langle k(\calA) \cdot |\calA| \rangle 
                   - \langle k(\calA) \rangle \langle |\calA|\rangle \right)$.
    \item $C_2 
    = L^{-d} \left(\langle \left[|\calA|/2 + k(\calA) \right]^2 \rangle 
                   - \langle |\calA|/2 + k(\calA) \rangle ^2\right) 
    =C_k +  C_{kb} + C_b/4$.
  \end{itemize}
    Here, the factor $1/2$ in the definition of $C_2$ arises from the fact
that the critical line $u_c(q)$ of 
    the RC model~(\ref{zrc}) on the square lattice is described by $u_c(q) = q^{1/2}$, which leads to a weight
    $q^{|\calA|/2+k(\calA)}$ for a subgraph $\calA$ along the critical line. 
In the canonical ensemble, the bond number $|\calA|$ is fixed, 
and $C_b$ and $C_{kb}$ reduce to $0$. 
\item {\it Magnetic quantities}: 
   \begin{itemize}
     \item The largest-cluster size $S_1=\langle \calC_1 \rangle$. 
     \item The cluster-size moments $S_\ell$ ($\ell>1$). 
Defining $\calS_\ell = \sum_i^k \calC_i^\ell \; ,$
we sampled $S_2 = \langle \calS_2 \rangle$ and $\langle 3 \calS_2^2 - 2\calS_4 \rangle$. 
  \end{itemize}
\item {\it Dimensionless quantities}: 
  \begin{itemize}
    \item Wrapping probabilities $R_1, R_b, R_x$, and $R_e$. 
      The probability $R_1$ counts the events that configuration $\calA$
      connects to itself 
      along the $x$ direction, but not along the $y$ direction;
      $R_b$ is for simultaneous wrapping in both directions; 
      $R_x$ is for the $x$ direction, irrespective of the $y$ direction; 
      $R_e$ is for wrapping in at least one direction.  
      They are related as $R_e = 2 R_1 + R_b$ 
      and $R_x = R_1 + R_b$, thus only two of them are independent~\cite{NZ,NZ1}.
    \item Universal ratio $Q_S= \langle \calC_1^2 \rangle/S_1^2$.
    \item Universal ratio $Q_m = \langle \calS_2 \rangle ^2 /
 \langle 3 \calS_2 ^2  -2 \calS_4 \rangle$.
  \end{itemize}
\end{enumerate}

\subsection{Finite-size scaling in the grand-canonical ensemble}
\label{expsca-grand}
The critical FSS of the sampled quantities can be obtained from derivatives 
of the free-energy density $f=- L^{-d} \ln \calZ_{\rm rc}$ ($\calZ_{\rm rc}$ is given by Eq.~(\ref{zrc})) 
with respect to the thermal scaling field $t$, the magnetic scaling field $h$, or the parameter $q$. 
In the grand-canonical ensemble, the FSS of $f(q,\,t,\,h,\,L)$ is expected to behave as
\begin{equation}
  f(q,\,t,\,h,\,L)
  = f_r(q,\,t,\,h)+ L^{-d} f_s(q,\,t L^{y_t},\, hL^{y_h},\,1) \; ,
\label{eq:grand-f}
\end{equation}
where higher-order scaling fields have been neglected, and $f_r$ and 
$f_s$ denote the regular and the singular part of the free-energy density, respectively.
The thermal scaling field $t$ is approximately proportional to $u-u_c$ in Eq.~(\ref{zrc}), 
where $u_c (q) =\sqrt{q}$ is the critical line of the $q$-state RC model on the square lattice. 

{\em FSS for energy-like quantities}. 
From the partition sum in Eq.~(\ref{zrc}) and Eq.~(\ref{eq:grand-f}), it can be derived 
that at criticality 
\begin{eqnarray}
  (-q) ({\rm d} f_c/{\rm d} q) 
    &=& L^{-d} \langle |\calA|/2 + k (\calA) \rangle 
    \equiv \rho_b (L) + \rho_k (L) 
     = \rho_{b,0} + \rho_{k,0}  + b L^{-d} \;, \label{eq:scaling-rho0} \\
  (-u/2) (\partial f/\partial u) 
    &=& 2^{-1} L^{-d} \langle |\calA| \rangle \equiv \rho_b (L)
     = \rho_{b,0} +  a L^{y_t-d} \;, \label{eq:scaling-rhob} \\
  (-q) (\partial f/\partial q)  
    &=& L^{-d} \langle k(\calA) \rangle \equiv  \rho_k (L)
     = \rho_{k,0} - a L^{y_t-d} + b L^{-d} \; ,
   \label{eq:scaling-rhok}
\end{eqnarray}
where $f_c = f(u=u_c)$ is the free-energy density along the critical line 
$u_c(q) = \sqrt{q}$.  The last equality in Eq.~(\ref{eq:scaling-rho0})
reflects the analyticity of $f_c$ (including the amplitude of its
finite-size dependence) along the critical line~\cite{KZ}.
Moreover, the correction 
amplitude $b$ is universal~\cite{ZFA, KZ}, which is also referred 
to as the excess cluster number for percolation. The term $\rho_{k,0}$ 
follows from the exact results for the critical Potts free energy on 
the square lattice \cite{Temperley-71}. An exact result is also available for the
triangular lattice \cite{BTA}.  The last equality in 
Eq.~(\ref{eq:scaling-rhob}) arises from Eq.~(\ref{eq:grand-f}), and $\rho_{b,0}$ 
accounts for the background contribution; the self-duality of the
square-lattice model yields $\rho_{b,0}(q)=1/2$.  In the $q \to 1$ limit, 
the amplitude $a$ vanishes as $a \sim q-1$. Therefore, one has for
critical percolation 
\begin{equation}
  \rho_b (L) = \rho_{b,0} \;, 
  \hspace{20mm} \rho_k(L) = \rho_{k,0} + bL^{-d} \; .
\label{eq:scaling-rhobk}
\end{equation}

{\em FSS for specific-heat-like quantities}. 
Similarly, one has the second derivatives (at $q=1$) as
\begin{eqnarray}
 -\left(q\frac{\rm d}{{\rm d}q}\right)^2 f_c& \equiv & C_2 = C_{2,0} + c L^{-d} \; , 
\label{eq:scaling-C2} \\
 -\left(u\frac{\partial}{\partial u}\right)^2  f & \equiv & C_b  = C_{b,0} \; , 
\label{eq:scaling-Cb} \\
 -\left(
        u\frac{\partial}{\partial u}\right) \left(q\frac{\partial}{\partial q}
  \right)  f 
 &\equiv& C_{kb} = C_{kb,0} + 2a' L^{y_t-d} + c' L^{-d} \; ,
\label{eq:scaling-Ckb} \\
 -\left(q\frac{\partial}{\partial q}\right)^2 f 
 &\equiv& C_k = C_{k,0} - 2a' L^{y_t-d} + c'' L^{-d} \; , 
\label{eq:scaling-Ck} 
\end{eqnarray}
where the last equalities in Eq.~(\ref{eq:scaling-Ckb}) and 
(\ref{eq:scaling-Ck}) arise from the fact that the parameter $a$ in Eqs.~(\ref{eq:scaling-rhob}) and (\ref{eq:scaling-rhok})
behaves as $a \approx a'(q-1)$. 
The analyticity of $f_c$ along the critical line~\cite{KZ} is also reflected 
in Eq.~(\ref{eq:scaling-C2}).  
It is also derived that the amplitude $c$ is a universal quantity~\cite{KZ}, called
the excess fluctuation of the number $|\calA|/2+k(\calA)$ for critical bond percolation. 

{\em FSS for magnetic quantities}.
The critical FSS of magnetic quantities can be obtained by differentiating
the free-energy density with respect to the magnetic scaling field $h$. 
It can be shown that at criticality,
\begin{equation}
S_1 (L) \sim L^{y_h}\, , \,\,\,\, 
S_2 (L) \sim L^{2 y_h}\, . 
\label{eq:scaling-mag}
\end{equation}

{\em Corrections to FSS}. 
At criticality, the asymptotic behavior of a quantity $O(L)$ 
is supposed to follow the form 
\begin{equation}
O(L)=L^{\psi} (O_0 + {\rm corrections}),
\label{eq-corr}
\end{equation}
where $\psi$ is the leading critical exponent and $O_0$ is the amplitude. 
For our observables in bond percolation, values of $\psi$ are given by 
the previous analysis ($\psi=0$ for dimensionless quantities). 
The FSS theory predicts several types of ``{\rm corrections}'' in Eq.~(\ref{eq-corr}) 
(see e.g. Refs.\cite{GOPS} and \cite{FDB} for a review and references therein). 
These terms include corrections from the irrelevant scaling fields; the leading one 
has exponent $y_i=-2$ for percolation.
A regular background term, which appears e.g. as $L^{-y_h}$ for $S_1$ and
$L^{d-2y_h}$ for $S_2$, contributes to the ``{\rm corrections}". 
These regular background terms are also present in the associated dimensionless ratios. 
For the energy-like quantities ($\rho_k$ and $\rho_b$) 
and the specific-heat-like quantities, $\psi$ is negative, thus the regular
background (which is $L^{-\psi}$ here) included in the ``{\rm corrections}'' in Eq.~(\ref{eq-corr}) is 
actually not a correction, instead it describes the leading behavior, which
is the thermodynamic limit of the sampled quantity. 

\section{Finite-size scaling in the canonical ensemble}
\label{expsca-cano}
As mentioned in the Introduction, 
the critical FSS for bond percolation in the canonical ensemble 
is due to the suppression of the fluctuation of occupied-bond density $\rho_b$. 
In this section, we demonstrate how the critical FSS is 
affected by such a suppression. 

Without the constraint, any configuration $\calA$ with bond number $|\calA|=N_b $ occurs 
with probability $p^{N_b} (1-p)^{(N_e-N_b)}$ ($N_e = |E|$). 
Thus, the value $O_g$ of an observable $O$ in the grand-canonical ensemble can be written as 
\begin{eqnarray}
O_g(p,L) 
&=& \sum_{N_b=0}^{N_e} p^{N_b}(1-p)^{N_e-N_b} \sum_{\calA:|\calA|=N_b} O(\calA) \; \nonumber \\
&=& \sum_{N_b=0}^{N_e} p^{N_b}(1-p)^{N_e-N_b} \frac{N_e!}{N_b! (N_e-N_b)!} O_c (\rho, L)  \; ,
\label{OgOc0}
\end{eqnarray}
where $\rho = N_b/N_e$ is the occupied-bond density and $O_c$ is obtained by averaging over all the configurations $\calA$
with $|\calA|=N_b$, of which the total number is $N_e!/N_b! (N_e-N_b)!$. Namely, $O_c$ is the canonical-ensemble value 
of observable $O$.
For a sufficiently large system, the binomial distribution in Eq.~(\ref{OgOc0}) is well approximated by a Gaussian distribution
\begin{eqnarray}
f(p, \rho, L) =\frac{1}{\sqrt{2 \pi} \sigma} e^{-(\rho-\arho)^2/{2\sigma^2}} \; ,
\,\,\,\, \sigma=\sqrt{{p(1-p)}/{N_e}} \; ,
\label{eq_Gaus}
\end{eqnarray}
where $\arho = p$ is the average bond density for occupation probability $p$, and the
variance $\sigma$ decreases as $\propto 1/ \sqrt{N_e} \propto L^{-d/2}$.
Thus, Eq.~(\ref{OgOc0}) can be re-written as
\begin{eqnarray}
\nonumber\\
O_g(p, L) &=& \int {\rm d} \rho \; f(p, \rho, L) O_c(\rho, L) \; .
\label{OgOc}
\end{eqnarray}
In principle, based on the known scaling behavior of $O_g$ in the grand-canonical ensemble,
the critical FSS of $O_c$ 
can be obtained by the inverse orthogonal transformation of Eq.~(\ref{OgOc}). 
Here, we take an approximation approach by Taylor expanding $O_c(\rho,L)$ in Eq.~(\ref{OgOc}) 
in powers of $\delta \rho=\rho-\arho$ about $\rho=\arho=p$, and term-by-term evaluation of
the integrals:
\begin{eqnarray}
 \int {\rm d} \rho \; f(p, \rho, L) O_c(\rho, L) 
&=& O_c(\arho, L) + O_c' \langle \delta \rho \rangle_f + O_c'' \langle (\delta \rho)^2/2 \rangle_f  + \ldots 
\label{OgOc1}
\end{eqnarray}
where the derivatives $O_c'$ and $O_c''$ are taken at $\arho$, and the average $ \langle \; \; \rangle_f$ is over 
the Gaussian distribution in Eq.~(\ref{eq_Gaus}). 
It is easily derived that $\langle \delta \rho \rangle_f =0 $ and $\langle (\delta \rho)^2 /2 \rangle_f = p (1-p)/2N_e$.
Combination of Eqs.~(\ref{OgOc}) and (\ref{OgOc1}) yields
\begin{equation}
  O_g(p, L) \simeq O_c(\arho, L) + B O_c'' L^{-d} \; ,
  \label{eq_central}
\end{equation}
with $B=p (1-p)/4$ a non-universal constant.

Near criticality $p_c$,  we denote $\Delta p = p-p_c$. Since  $\arho=p$, we can equivalently
denote the distance to the percolation threshold as $\Delta \arho = \arho - \arho_c$ with
$\arho_c=p_c$ and $\Delta \arho = \Delta p$. Thus
\begin{eqnarray}
  O_g(\Delta p, L) &=& O_{g,r} (\Delta p) + L^{\psi} [O_{g,s} (L^{y_t} \Delta p ) + O_{g,i} L^{y_i}],
\label{Og}\\
O_c(\Delta \arho, L) &=& O_{c,r} (\Delta p) + L^{\psi'} [O_{c,s} (L^{y_{\rho}} \Delta p )+ B_{\rm can} L^{\yc} + O_{c,i} L^{y_i}] \; ,
\label{Oc}
\end{eqnarray}
where new finite-size corrections with exponent $\yc$ are allowed in Eq.~(\ref{Oc}). 
It is known from Eq.~(\ref{Oc}) that $O_c''(\arho_c,L) = O_{c,r}''+ L^{\psi'+2y_\rho} O_{c,s}''$.
Taylor-expanding the r.h.s of Eqs.~(\ref{Og}-\ref{Oc}) and substituting them in Eq.~(\ref{eq_central}) give
\begin{eqnarray}
  O_{g,r} (0) &+& L^{\psi}  [O_{g,s} (0) + O_{g,s}' L^{y_t}    \Delta p  + O_{g,i} L^{y_i} +\ldots ] =   \label{eq_OgOc0} \\
  O_{c,r} (0) &+& L^{\psi'} [O_{c,s} (0) + O_{c,s}' L^{y_\rho} \Delta p +  O_{c,i} L^{y_i} + B O_{c,s}'' L^{2y_\rho -d} + B_{\rm can} L^{\yc} +
  B O_{c,r}'' L^{-d-\psi'} + \ldots] \nonumber 
\end{eqnarray}

Assuming that $2y_\rho -d <0$, the term with $O_{c,s}''$ acts as a finite-size correction. 
By comparing the scaling behavior of the l.h.s and the r.h.s. of Eq.~(\ref{eq_OgOc0}), we obtain
\begin{enumerate}
  \item $\psi' = \psi$ and $y_\rho = y_t$, 
  \item $O_{c,r} (0) = O_{g,r} (0)$, $O_{c,s} (0) = O_{g,s} (0)$ (for $\psi \neq -d$), and $O_{g,s}' = O_{c,s}'$ etc. 
        In other words, the scaling functions in Eqs.~(\ref{Og}) and (\ref{Oc}) are identical up to some correction terms.
  \item new correction terms appear in the canonical ensemble, and they have exponents $2y_t-d$ and $-d-\psi$ (if $-d-\psi<0$).
        This is because the leading finite-size correction in the l.h.s of Eq.~(\ref{eq_OgOc0}) is described by an exponent $y_i=-2$,
	and thus the terms with $L^{2y_\rho -d}=L^{2y_t-d}$ and $ L^{-d-\psi'}=L^{-d-\psi}$ 
	have to be cancelled by the newly included correction terms 
	with $L^{\yc}$. It can be shown that corrections with exponent $n(2y_t-d)$ with $n=2,3,\ldots$ can also occur 
	if higher-order terms are kept in the above Taylor expansions.
  \item for the case of $\psi = -d$, $O_{c,s} (0) = O_{g,s} (0)$ does not hold. 
        Instead, one has $ O_{g,s} (0) = O_{c,s} (0)+ B O_{c,r}''$. 
	This predicts that the excess cluster number $b$ in Eq.~(\ref{eq:scaling-rhobk}) is changed by the constraint;
	further, since $B$ and $\rho''_{c,r}$ are non-universal, the canonical-ensemble value of $b$ is no longer universal.
\end{enumerate}
 Note that the assumption $2y_\rho -d <0$ holds since $2y_t-d<0$ for percolation in two and higher dimensions. 

 Finally, we mention that although the constraint is ``energy-like'', the derivation 
 applies to both the energy-like and the magnetic observables.

\section{Numerical results} 
\label{num_results}
To examine the above theoretical derivations, we carry out Monte Carlo simulations 
for the critical bond percolation on the square lattice, with a fixed bond number 
$N_b = p_c |E| =L^2$, and for $21$ sizes in range $4 \leq L \leq 4000$. The number of samples was about
$10^8$ for $L \le 480$, $5\times 10^7 $ for $L = 800$, $2.5\times 10^7 $ 
for $L = 1600$, and $10^6$ for $L = 4000$.
For the purpose of comparison, additional simulations were also carried out
in the grand-canonical ensemble.

\subsection{Evidence for the correction exponent $y_{\rm can}=2y_t-d$}
\label{sec-ycan}
\begin{figure}
\includegraphics[width=10.0cm]{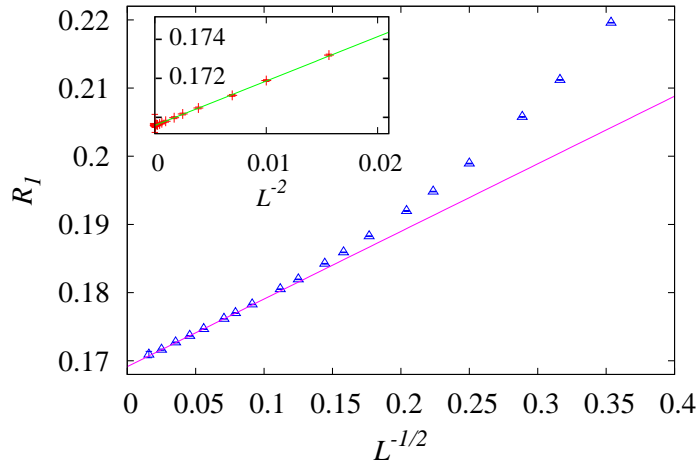}
\caption{(Color online). 
Plot of the wrapping probability $R_1$ versus $L^{-1/2}$ in the
canonical ensemble (main plot). 
The inset shows $R_1$ versus $L^{-2}$ in the grand-canonical ensemble.
The straight lines represent the leading finite-size dependence as 
obtained from the fit. 
The error bars of the data are smaller than the size of the data points.} 
\label{figure-R1}
\end{figure}

We first examine the critical FSS of the wrapping probabilities. 
In the grand-canonical ensemble, the finite-size corrections arise only from the irrelevant scaling fields 
for which the leading exponent $y_i=-2$. This is illustrated in the inset of Fig.~\ref{figure-R1}.
In the canonical ensemble, the existence of the newly induced correction exponent $y_{\rm can}=2y_t-d=-1/2$
is clearly demonstrated by the approximately linear behavior for large $L$ in Fig.~\ref{figure-R1},
where $R_1$ is plotted versus $L^{-1/2}$.

According to the least-squares criterion, we fitted the data for these
wrapping probabilities by  
\begin{equation}
O=O_0+B_1 L^{\yc}+B_2 L^{2 \yc} + B_i L^{y_i},
\label{fit_R}
\end{equation}
where $O_0$ represents the universal value for $L \rightarrow \infty$,
and the correction exponent $y_i$ is fixed at $-2$. 
As a precaution against higher-order correction terms not
included in the fit formula, the data points for small $L<L_{\rm min}$ 
were gradually excluded to see how the residual $\chi^2$ changes with 
respect to $L_{\rm min}$. In general, we use results of the fit 
corresponding to a $L_{\min}$ for which the quality of fit is 
reasonable, and for which subsequent increases of $L_{\rm min}$ do not   
cause the $\chi^2$ value to drop by vastly more than one unit per degree
of freedom.
In practice, the word ``reasonable" means here that $\chi^2$ is less than
or close to the number of degrees of freedom, and that the fitted parameters
become stable. 

\begin{table}
\caption{Fit results for wrapping probabilities. 
The exact values of $O_0$~\cite{Pinson, ZLK}, and $y_i=-2$ are used here.
The abbreviations ``GE" and ``CE" stand for grand- and canonical ensemble, respectively. 
Entries ``-" indicate the absence of a result,
and the numbers without error bars are fixed in the fits.
Error margins are quoted as two times of the statistical errors in the fits.}
\label{fit_dimensionless}
\begin{center}
    \begin{tabular}{c|l|l|l|l|l|l|l|l}
    \hline 
    \hline 
	\multicolumn{9}{c}{Fit using $O = O_0 + B_1 L^{\yc} + B_2 L^{2 \yc} + B_i L^{y_i}$} \\
    \hline 
\multicolumn{1}{c}{}
& & $O_0$ & $B_1$ & $\yc$ & $B_2$ & $B_i$ & $y_i$ & $L_{\rm min}$ \\

    \hline 
& GE & $0.169\,415\,435$ & $-$ & $-$ & $-$ & $0.23(1)$ & $-2$ & $16$ \\
&    & $0.169\,415\,435$ & $-$ & $-$ & $-$ & $0.24(3)$ & $-2.02(4)$ & $8$ \\
    \cline{2-9} 
$R_1$ 
&    &$0.169\,5(4)$ &$0.11(2)$ & $-0.53(5)$ & $-$ & $-$ & $-$ & $160$ \\
& CE & $0.169\,415\,435$ & $0.091\,(8)$ & $-0.502\,(13)$ & $0.08(2)$ & $0.56(7)$ & $-2$ & $12$ \\
&    & $0.169\,415\,435$ & $0.089\,2(5)$ & $-1/2$ & $0.082\,(4)$ & $0.55(3)$ & $-2$ & $12$ \\

    \hline 
& GE & $0.351\,642\,855$ & $-$ & $-$ & $-$ & $-0.254\,(11)$ & $-2$ & $12$ \\
&    & $0.351\,642\,855$ & $-$ & $-$ & $-$ & $-0.30(5)$ & $-2.06(8)$ & $8$ \\
    \cline{2-9} 
$R_b$ 
&    & $0.351\,5(6)$ & $-0.10(3)$ & $-0.55(7)$ & $-$ & $-$ & $-$ & $120$ \\ 
& CE & $0.351\,642\,855$ & $-0.075\,(11)$ & $-0.50(2)$ & $-0.07(2)$ & $-0.53(5)$ & $-2$ & $8$ \\ 
&    & $0.351\,642\,855$ & $-0.075\,8(8)$ & $-1/2$ & $-0.066\,(5)$ & $-0.53(3)$ & $-2$ & $8$ \\ 

    \hline 
& GE & $0.521\,058\,290$ & $-$ & $-$ & $-$ & $-0.021\,(16)$ & $-2$ & $16$ \\
&    & $0.521\,058\,290$ & $-$ & $-$ & $-$ & $-0.1(1)$ & $-2.4(6)$ & $8$ \\
    \cline{2-9} 
$R_x$ 
&    & $0.521\,1(2)$ & $0.020\,(4)$ & $-0.56(8)$ & $-$ & $-$ & $-$ & $32$ \\ 
&    & $0.521\,058\,290$ & $0.018\,(7)$ & $-0.54(8)$ & $-$ & $-$ & $-$ &$120$ \\
& CE & $0.521\,058\,290$ & $0.014\,(9)$ & $-0.51(9)$ & $0.016(13)$ & $-$ & $-$ & $16$ \\
&    & $0.521\,058\,290$ & $0.016\,(12)$ & $-0.52(11)$ & $0.01(3)$ & $0.03(8)$ & $-2$ & $10$ \\
&    & $0.521\,058\,290$ & $0.013\,4(8)$ & $-1/2$ & $0.016\,(6)$ & $0.02(4)$ & $-2$ & $10$ \\

    \hline 
& GE & $0.690\,473\,725$ & $-$ & $-$ & $-$ & $0.22(2)$ & $-2$ & $16$ \\
&    & $0.690\,473\,725$ & $-$ & $-$ & $-$ & $0.18\,(4)$ & $-2.0(1)$ & $8$ \\
    \cline{2-9} 
$R_e$ 
&    & $0.690\,5(6)$ &$0.13(3)$ & $-0.53(6)$ & $-$ & $-$ & $-$ & $120$ \\ 
&    & $0.690\,473\,725$ & $0.13(3)$ & $-0.53(3)$ & $-$ & $0(7)$ & $-2$ & $120$ \\
& CE & $0.690\,473\,725$ & $0.10(3)$ & $-0.50(3)$ & $0.13(4)$ & $-$ & $-$ & $40$ \\
&    & $0.690\,473\,725$ & $0.12(3)$ & $-0.52(3)$ & $0.05(9)$ & $0.9(5)$ & $-2$ & $20$ \\
&    & $0.690\,473\,725$ & $0.102(2)$ & $-1/2$ & $0.10(2)$ & $0.5(3)$ & $-2$ & $24$ \\

    \hline 
    \hline 
    \end{tabular}
\end{center}
\end{table}

\begin{figure}
\includegraphics[width=10.0cm]{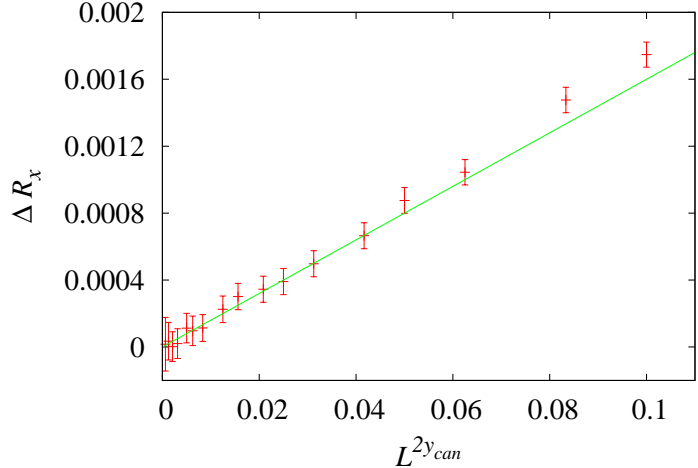}
\caption{(Color online). 
Plot of $\Delta R_x=R_x-O_0-B_1 L^{\yc}$ versus $L^{2\yc}$ for 
the canonical wrapping probability $R_x$. 
The straight line represents the leading finite-size dependence of
$\Delta R_x$, i.e. $B_2 L^{2\yc}$. 
Parameter $O_0$ and $\yc$ are fixed at the exact value and $-1/2$ respectively, 
and the amplitudes $B_1$, $B_2$ are obtained from the fit.}
\label{figure-dRx}
\end{figure}

The results are shown in Table~\ref{fit_dimensionless}. The error margins
are quoted as two times of the statistical errors in the fits, which also applies 
to other tables, in order to account for possible systematic errors. 
The universal values $O_0$ can be exactly obtained~\cite{Pinson, ZLK}.
In the grand-canonical ensemble, these exact values of $O_0$ were used and the amplitudes $B_1$ and $B_2$ were set at $0$,
so that the fit formula in Eq.~(\ref{fit_R}) has only a single free parameter.
Indeed, such a simple formula can well describe the data with $L \geq 16$ for 
all the wrapping probabilities ($R_1$, $R_b$, 
$R_x$, and $R_e$); further, for $R_x$ the amplitude of $B_i$ is very small.
As expected, including terms with $B_1$ and/or $B_2$ only yields messy information and 
does not improve the quality of the fits. 
In the canonical ensemble, we also fitted the data by Eq.~(\ref{fit_R}) with a single correction term $B_1 L^{\yc}$ 
($B_2=B_i=0$). As shown in Tab.~\ref{fit_dimensionless}, 
the data up to sufficiently large $L_{\rm min}$ (except $R_x$) have to be 
discarded for a reasonable residual $\chi^2$.
Nevertheless, we find that (i), the correction term has an exponent close to $-1/2$, 
and (ii), the estimates of $O_0$ agree well with the exact values.
A better description of the $R$ data can be obtained by including correction terms with $B_i$ and/or $B_2$;
for simplicity the exact values of $O_0$ are used. 
As a result, the estimates of $\yc$ become more accurate, and are in good agreement with the prediction $\yc=-1/2$.
We note that the correction coefficient $B_i$ takes different values in the grand-canonical and the canonical ensemble.
This is also in agreement with the theoretical expectation. 
As predicted in Sec. III, in addition to those with exponent $n(2y_t-d)$ $(n=1,2,\ldots)$, 
correction terms with exponent $-d-\psi$ can also exist in the canonical ensemble. 
For the wrapping probabilities, one has $\psi=0$ and thus such a correction term 
has the same exponent as $B_i L^{y_i}$.
Finally, in order to illustrate the term with $L^{2\yc}(=L^{-1})$,
we plot $\Delta R_x = R_x - O_0 - B_1 L^{\yc}$ versus $L^{2\yc}$ as in Fig.~\ref{figure-dRx}. 

In order to demonstrate that the correction exponent for the grand-canonical wrapping probabilities is indeed $y_i=-2$, 
we also fitted the data by Eq.~(\ref{fit_R}) with $y_i$ being free, $O_0$ being fixed at the exact values and $B_1=B_2=0$. 
The fit results are also included in Table~\ref{fit_dimensionless}.
The result is   $-2.02(4)$, $-2.06(8)$, $-2.4(6)$ and
$-2.0(1)$, for $R_1$, $R_b$, $R_x$ and $R_e$, respectively. 
These results are in compatible with the existing numerical result 
$y_i=-2.003\,(5)$~\cite{NZ1} for $R_1$, and agree well with the exact value $y_i=-2$.
It is interesting to mention that although correction of form $L^{-2} (1+A \log L)$ ($A$ is a constant) has been 
observed for other observables in the square-lattice bond percolation~\cite{FDB}, there is no evidence for 
such a logarithmic factor in the finite-size corrections for the wrapping probabilities.

\begin{table}
\caption{Fit results for $S_1$, $S_2$, and the universal ratios $Q_S$ and $Q_m$.}
\label{fit_mag}
\begin{center}
    \begin{tabular}{c|l|l|l|l|l|l|l}
    \hline 
    \hline 
	\multicolumn{8}{c}{Fit using
$O = L^{\psi} (O_0 + B_1 L^{\yc} + B_r L^{y_r} + B_i L^{y_i})$} \\
    \hline 
\multicolumn{1}{c}{}
& & $O_0$ & $B_1$ & $\yc$ & $B_r$ & $B_i$  & $L_{\rm min}$ \\ 
    \hline 
$S_1$ 
& GE & $0.984\,48(3)$ & $-$ & $-$ & $1.7(13)$ & $-2(2)$ & $32$ \\
    \cline{2-8} 
& CE & $0.984\,3(4)$ & $0.066\,(14)$ & $-0.50(5)$ & $6(5)$ & $-7(6)$ & $24$ \\
    \hline 
$S_2$ 
& GE & $1.035\,97(4)$ & $-$ & $-$ & $1.7(7)$ & $-3.0(14)$ & $32$ \\ 
    \cline{2-8} 
& CE & $1.035\,8(5)$ & $0.05(3)$ & $-0.51(12)$ & $5(3)$ & $-7(5)$ & $24$ \\
    \hline 
$Q_{S}$ 
& GE & $0.960\,17(1)$ & $-$ & $-$ & $1.4(8)$ & $-1.8(12)$ & $40$ \\
    \cline{2-8} 
& CE & $0.960\,0(3)$ & $0.072\,(8)$ & $-0.48(4)$ & $-$ & $-$ & $200$ \\
&    & $0.960\,03(13)$ & $0.075\,(5)$ & $-0.49(2)$ & $-0.3(16)$ & $0(2)$ & $24$ \\
    \hline 
$Q_m$ 
& GE & $0.870\,56(2)$ & $-$ & $-$ & $-1.1(3)$ & $1.92(6)$ & $32$ \\
    \cline{2-8} 
& CE & $0.870\,4(2)$ & $0.138\,(7)$ & $-0.491\,(13)$ & $-1.1(6)$ & $1.3(9)$ & $16$ \\
    \hline 
    \hline 
    \end{tabular}
\end{center}
\end{table}

For magnetic quantities, we consider the largest-cluster size $S_1$ and the second cluster-size moment $S_2$, 
as well as the dimensionless ratios. 
We fitted the data by
\begin{equation}
O=L^{\psi} (O_0+B_1 L^{\yc}+B_r L^{y_r} + B_i L^{y_i}),
\label{fit_eq_mag}
\end{equation}
with $y_i=-2$ and $\psi$ being fixed at the exactly known value for the two-dimensional percolation.
The term with $L^{y_r}$ is the correction
from the regular background, with $y_r=-y_h = -91/48$ for $S_1$ and $Q_S$, and $y_r=d-2y_h=-43/24$ for $S_2$ and $Q_m$. 
The results are shown in Table~\ref{fit_mag}.
Again, we find that (i), the values of $O_0$ remain unchanged in the canonical ensemble, 
irrespective of whether they are universal (for $Q_S$ and $Q_m$) or non-universal (for $S_1$ and $S_2$),
and (ii), new correction terms with an exponent $\yc = -1/2$ are introduced. 
The estimate $Q_{m,0}=0.870\,56(2)$ in the grand-canonical ensemble agrees with 
the existing result $0.870\,53(2)$~\cite{DB1}.

\begin{table}
\caption{Fit results for the cluster-number density $\rho_k$ as determined
for the bond (BP) and site (SP) percolation problem. 
Some fits make use of the exact results 
$\rho^{\rm bond}_{k,0}=0.098\,076\,211$~\cite{Temperley-71} 
and $b_g=0.883\,576\,308$~\cite{KZ}.}
\label{fit_res_rhok}
\begin{center}
\begin{tabular}{c|l|l|l|l|l|l|l}
    \hline 
    \hline 
	\multicolumn{8}{c}{Fit using
 $\rho_k =\rho_{k,0}+L^{-2} (b + B_1 L^{\yc} + B_i L^{y_i})$} \\
    \hline 
\multicolumn{1}{c}{} 
& & $\rho_{k,0}$ & $b$ & $B_1$ & $\yc$ & $B_i$  & $L_{\rm min}$ \\

    \hline 
& GE & $0.098\,076\,22(7)$ & $0.883\,6(6)$ & $-$ & $-$ & $0.18(3)$ & $6$ \\
BP 
&   & $0.098\,076\,211$ & $0.883\,576\,308$ & $ $ & $ $ & $0.178\,(14)$ & $6$ \\
    \cline{2-8} 
&    & $0.098\,076\,21(4)$ & $0.23(2)$ & $0.57(5)$ & $-0.47(6)$ & $0.0(4)$ & $8$ \\
& CE & $0.098\,076\,211$ & $0.226\,(8)$ & $0.564\,(12)$ & $-0.47(3)$ & $-$ & $10$ \\
&    & $0.098\,076\,211$ & $0.233\,(4)$ & $0.60(2)$ & $-1/2$ & $-$ & $24$ \\
\hline

& GE & $0.027\,598\,03(2)$ & $0.883\,5(2)$ & $-$ & $-$ & $-0.17(4)$ & $16$ \\
SP 
&    & $0.027\,598\,02(2)$ & $0.883\,576\,308$ & $-$ & $-$ & $-0.22(3)$ & $20$ \\
    \cline{2-8}
& CE & $0.027\,598\,00(4)$ & $0.41(2)$ & $0.64(7)$ & $-0.46(7)$ & $-$ & $24$ \\
&    & $0.027\,597\,99 (4)$ & $0.417\,(2)$ & $0.687\,(11)$ & $-1/2$ & $-$ & $24$ \\
    \hline 
    \hline 
\end{tabular}
\end{center}
\end{table}

\subsection{Change of the universal excess cluster number}
\label{sec-uni}

For energy-like quantities, we consider the cluster-number density $\rho_k$, and 
analyze the data by
\begin{equation}
\rho_k =\rho_{k,0}+L^{-2} (b + B_1 L^{\yc} + B_i L^{y_i}),
\label{fit_rhok}
\end{equation}
with $y_i=-2$ being fixed. 
The background term can be exactly obtained as $\rho^{\rm bond}_{k,0}=(3\sqrt{3}-5)/2$ 
for the critical square-lattice bond percolation~\cite{Temperley-71}.
In the grand-canonical ensemble, the  excess cluster number $b$ 
is known to be universal and the value has been obtained as $b_g=0.883\,576\,308$~\cite{KZ},
with subscript $g$ for the grand-canonical ensemble.

\begin{figure}
\includegraphics[width=10.0cm]{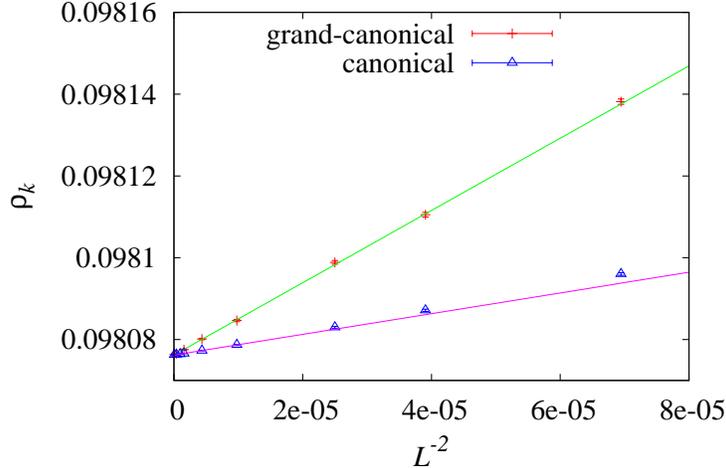}
\caption{(Color online).
Plot of the cluster-number density $\rho_k$ versus $L^{-2}$ for the bond
percolation model. The straight lines represent the leading finite-size 
dependence of $\rho_k$ as obtained from the fit.} 
\label{figure-rhok}
\end{figure}

The fit results are given in Table~\ref{fit_res_rhok}. 
As in the above subsection, the correction term with $\yc=-1/2$ is observed in the canonical ensemble, 
and the background contribution $\rho_{k,0}$ remains unchanged.
However, the excess cluster number is now $b_c = 0.233 (4)$, clearly different from the universal value 
$b_g=0.883\,576\,308$~\cite{KZ} in the grand-canonical ensemble. 
This indeed confirms the theoretical prediction for the case of $\psi=-2$ in Sec.~\ref{expsca-cano}.
In other words, one expects that $b_g = b_c + B \rho''_{k,r}$, 
where $B=p(1-p)/4$ is a constant and $\rho''_{k,r}$ is the second derivative $\rho''_{k,r}$ 
of the regular part of $\rho_k$ with respect to the bond density. 
Since both $B$ and $\rho''_{k,r}$ are non-universal, the canonical-ensemble value $b_c$ is {\it no longer} universal. 
As an illustration, we plot $\rho_k$ versus $L^{-2}$ in Fig.~\ref{figure-rhok},
where the difference of $b_g$ and $b_c$ is reflected by the different slopes of the data lines. 

In order to examine the {\it ``non-universal"} nature of the excess cluster number $b_c$ in the
canonical ensemble, we also performed simulations of square-lattice
site-percolation problem at the percolation threshold 
$p^{\rm site}_{c}=0.592\,746\,02$~\cite{Lee, FDB, Ziff11}. 
In the grand-canonical ensemble, the simulations
 used 15 system sizes in range $4 \le L \le 512$,
and the results of the fits by Eq.~(\ref{fit_rhok}) are given in Tab.~\ref{fit_res_rhok}. 
The estimate of $b_g=0.883\,5(2)$ agrees well with the universal value $0.883\,576\,308$~\cite{KZ}. 

In the canonical ensemble, the total number of occupied sites $N_s (L)=p^{\rm site}_{c} L^2$ is
fixed. However, $N_s (L)$ is not an integer, and thus the actual simulations were carried out 
for the total occupied-site number $[N_s (L)]$ and $[N_s (L)]+1$, with $[ \; ]$ for the floor integer.
The Monte Carlo results at  $N_s (L)$ were then obtained by linear interpolation.
The simulations used $16$  system sizes $L$ in range $4 \leq L \leq 1024$.
The results of fits using Eq.~(\ref{fit_rhok}) are also given in Tab.~\ref{fit_res_rhok}.
As expected, one has a correction term with exponent $-1/2$ and $b_c^{\rm (site)} [= 0.417(2)] \neq b_g$. 
The {\it ``non-universal''} property of $b_c$ is demonstrated by the fact $b_c^{\rm (site)} \neq b_c^{\rm (bond)}$. 

In addition, from our fits, one gets 
$\rho^{\rm site} _{k,0} = 0.027\,598\,00(5)$, 
which is in good agreement with the existing result $0.027\,598\,1(3)$~\cite{ZFA},
and reduces the error margin significantly. 

We also study the FSS of the specific-heat-like quantity $C_2$ for the bond percolation model and find 
that due to the limited precision, the data for $L \geq 8$ are well described 
by $C_2 (L) = C_{2,0} + c L^{-2}$. 
As already mentioned in Section \ref{simul}, the excess fluctuation $c$ is universal.
In the grand-canonical ensemble, the fit yields $C_{2,0}=0.039\,446\,(4)$ and $c_g = 0.105 \, 5(7)$;
the former is consistent with the existing result $C_{2,0}=0.039\,44(4)$~\cite{DY}, 
the latter agrees well with the exactly known value $c=0.105\,436\,634$~\cite{KZ}.
In the canonical ensemble, the result is  $C_{2,0}=0.039\,446\,(6)$ and $c_c=-0.33 (5)$. 
It is interesting to observe that not only the magnitude of the excess fluctuation $c$ is modified, 
but that also its sign has changed.

\subsection{Universality of scaling functions}
\begin{figure}
\includegraphics[width=10.0cm]{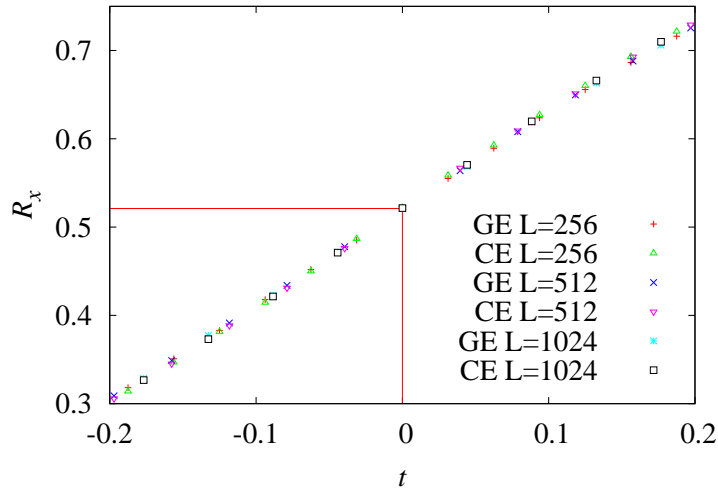}
\caption{(Color online). Plot of $R_x$ versus $t$ for bond percolation.
Parameter $t$ represents $L^{y_t}(p-p_c)$ in the grand-canonical ensemble (GE), 
and $L^{y_t} (\arho-\arho_c)$ in the canonical ensemble (CE). 
The excellent collapse demonstrates the universality of scaling functions
near the critical point.} 
\label{figure-Rxvsp}
\end{figure}

In Sec.~\ref{expsca-cano}, we state that the scaling functions in Eqs.~(\ref{Og})
and (\ref{Oc}) are identical up to some correction terms, which reflects the universality
of the functions. In order to demonstrate this, we carried out simulations 
near the critical point. We plot the grand-canonical $R_x$ versus $L^{y_t} (p-p_c)$ 
and the canonical $R_x$ versus $L^{y_t} (\arho-\arho_c)$ as in Fig.~\ref{figure-Rxvsp}. 
As expected, data points in both the ensembles nicely collapse to the same curve. 

\section{Discussion}
\label{discus}
We derive the critical finite-size scaling behavior of percolation under the constraint
that the total number of occupied bonds/sites is fixed, and confirm theoretical 
predictions for the two-dimensional percolation by means of Monte Carlo simulation in two dimensions.
In particular, it is found that with the constraint, new finite-size corrections 
with exponent $n(2y_t-d)$ $(n=1,2,\ldots)$ are induced and the excess cluster number becomes non-universal.
We note that our theory in Sec.~\ref{expsca-cano} can be used to explain the observed correction
exponent $\approx -0.5$ for the canonical wrapping probabilities
in simulating the two-dimensional percolation by the Newman-Ziff algorithm~\cite{NZ, NZ1}.
The predictions should be valid in any dimension with $d \geq 2$.
We believe that this work provides an additional useful reference for percolation---a
pedagogical system in the field of statistical mechanics. Furthermore, our work can
help to understand the critical finite-size-scaling properties of other
statistical systems in the canonical ensemble with a fixed total number of 
particles, which is the usual situation during experiments.

\acknowledgments
This work was supported by NSFC (under grant numbers 10975127, 91024026 and
11275185), CAS, and Fundamental Research Funds for the Central Universities 
No.\ 2340000034. We would like to thank R. M. Ziff for helpful comments.  
H. Hu is grateful for the hospitality of the Instituut Lorentz 
of Leiden University.

\end{document}